\documentclass{iau} 
\usepackage{graphicx}

\title[VLASS Variable Radio AGN] 
{Variable Radio AGN at High Redshift Identified in the VLA Sky Survey}

\author[Nyland et al.]{Kristina Nyland$^1$, Dillon Dong$^2$, Pallavi Patil$^{3,4}$, Mark Lacy$^4$, Amy Kimball$^5$, Gregg Hallinan$^2$, Sumit Sarbadhicary$^6$, Emil Polisensky$^7$, Namir Kassim$^7$, Wendy Peters$^7$, Tracy Clarke$^7$, Dipanjan Mukherjee$^8$, Sjoert van Velzen$^{9,10}$, Vivienne Baldassare$^{11,12}$}

\affiliation{
$^1$National Research Council fellow, resident at NRL, Washington, DC, USA; 
$^2$California Institute of Technology, Pasadena, CA, USA;
$^3$University of Virginia, Charlottesville, VA, USA; 
$^4$NRAO, Charlottesville, VA, USA; 
$^5$NRAO, Socorro, NM, USA; 
$^6$Michigan State University, Lansing, MI, USA; 
$^7$NRL, Washington, DC, USA; 
$^8$IUCAA, Pune, India; 
$^9$New York University, New York, NY, USA; 
$^{10}$University of Maryland, College Park, MD, USA
$^{11}$Yale University, New Haven, CT, USA
$^{12}$Einstein Fellow}

\pubyear{2020}
\volume{359}  
\jname{Galaxy evolution and feedback across different environments}
\editors{T. Storchi-Bergmann, R. Overzier, W. Forman \& R. Riffel, eds.}
\begin{document}

\maketitle

\begin{abstract}
As part of an on-going study of radio transients in Epoch 1 (2017--2019) of the Very Large Array Sky Survey (VLASS), we have discovered a sample of $0.2<z<3.2$ active galactic nuclei (AGN) selected in the optical/infrared that have recently brightened dramatically in the radio.  These sources would have previously been classified as radio-quiet based on upper limits from the Faint Images of the Radio Sky at Twenty-centimeters (FIRST; 1993-2011) survey;  however, they are now consistent with radio-loud quasars.  We present a quasi-simultaneous, multi-band (1--18 GHz) VLA follow-up campaign of our sample of AGN with extreme radio variability.  We conclude that the radio properties are most consistent with AGN that have recently launched jets within the past few decades, potentially making them among the youngest radio AGN known.  
\end{abstract}

\firstsection 

\section{Introduction}
The slow (timescales of seconds to years) radio transient population is dominated by active galactic nuclei (AGN), the majority of which are associated with variability (\cite[Mooley et al.\ 2016]{mooley+16}).   
AGN variability in the radio may arise from extrinsic propagation effects (e.g.\ interstellar scattering) or intrinsic mechanisms directly related to the AGN itself (e.g. \cite[Thyagarajan et al.\ 2011]{thyagarajan+11}). 
Whether driven by extrinsic or intrinsic effects, variable radio AGN beyond the low-$z$ universe are inherently compact in nature.  
AGN with compact, sub-galactic jets hosted by gas-rich galaxies, especially at 1$\lesssim z \lesssim 3$, are an important, yet still poorly studied, 
class of objects for understanding the link between jet-ISM feedback and galaxy  evolution.  
In particular, {\it the prevalence of compact radio jets as a function of redshift and host galaxy properties remains unknown.}  
AGN with compact jets are challenging to identify in single-epoch radio surveys due to observational limitations (e.g. resolution and sensitivity), as well as inherent limitations due to 
variability.  Recently, multi-epoch radio surveys have begun to make major advancements identifying AGN with compact jets on the basis of radio variability.  

\begin{figure}[t!]
\begin{center}
\includegraphics[clip=true, trim=0.25cm 7.85cm 0.2cm 6.25cm, width=0.95\textwidth]{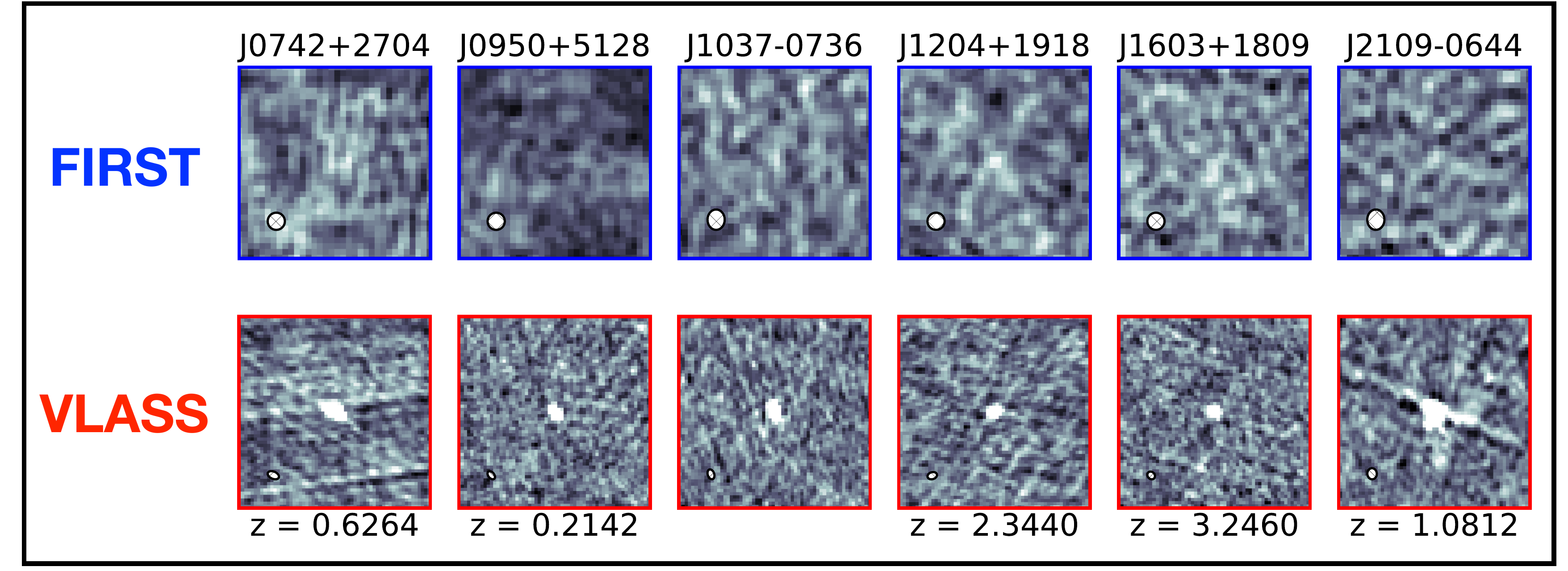} 
\caption{Cut-out images (1$^{\prime} \times 1^{\prime}$) from FIRST (1.4~GHz; 1993--2011) and VLASS Epoch 1 (3~GHz; 2017--2019) of a subset of our sample highlighting their variability on timescales of 1--2 decades. The synthesized beam is shown in the lower left corner of each cut-out image.  
} 
\label{fig:image_example}
\end{center}
\end{figure}

\section{Sample}
As part of our ongoing search for radio transients in Epoch~1 (2017-2019) of the Very Large Array Sky Survey (VLASS; \cite[Lacy et al. 2020]{lacy+20}), we have identified a preliminary sample of $\sim$2000 candidate ``transients" that are compact and $>$1~mJy in VLASS, but below the 5$\sigma$ detection threshold ($<0.75$~mJy) of the 1.4~GHz FIRST survey (\cite[Becker et al. 1995]{becker+95}) observed 1--2 decades earlier.  
VLASS is a synoptic survey at $S$-band (2--4~GHz) of the entire northern sky visible to the VLA ($\delta > -40^{\circ}$). 
A unique feature of VLASS is its synoptic design consisting of 3 epochs with a cadence of 32 months.  

{\underline{\it Selection of Variable Radio AGN}}. 
We considered optical and IR AGN selection diagnostics to capture both the unobscured and obscured AGN populations.  In the optical, we searched for spectroscopically verified quasars from the Sloan Digital Sky Survey (SDSS; \cite[York et al. 2000]{york+00}) DR14 (\cite[Paris et al. 2018]{paris+18}) and found 52 candidates 
within $1^{\prime \prime}$ of the VLASS position.  Using data from the Widefield Infrared Survey Explorer (WISE; \cite[Wright et al. 2010]{wright+10}) and the AGN diagnostic criteria from \cite[Assef et al. (2018)]{assef+18}, we found 144 obscured AGN with variable radio emission. We further required VLASS fluxes $>$3~mJy to rule-out steady but optically-thick sources with spectral indices up to $\alpha=2.5$ (where $S \sim \nu^{\alpha}$). 
Our final sample contains 26 radio-variable AGN in the 25\% of VLASS Epoch~1 analyzed thus far. 

{\underline{\it Source Properties}}. 
Spectroscopic redshifts ($0.2 \lesssim z \lesssim 3.2$) are available for the optically-selected subset of our sample (13/26 sources). These sources are classified as broad-line quasars and 
have radio luminosities of $\log(L_{3\,\rm GHz}/{\rm erg \, s^{-1}}) \approx40-42$.  \cite[Shen et al. 2011]{shen+11} report bolometric luminosities of $\log(L_{\rm bol}/{\rm erg \, s^{-1}}) \approx 45.2 - 46.8$ and virial supermassive black hole (SMBH) masses of $\log(M/{\rm M}_{\odot}) \approx 8.0-9.7$.  We are in the process of obtaining spectroscopic redshifts for the IR-selected, obscured AGN portion of our sample.  We note that some of the IR-selected AGN 
have extreme colors in WISE that are consistent with heavily obscured, hyperluminous quasars (\cite[Lonsdale et al. 2015]{lonsdale+15}, \cite[Patil et al. 2020]{patil+20}). 
Example cut-out images illustrating the remarkable increase in radio flux between FIRST and VLASS are shown in  Fig.\,\ref{fig:image_example}.

\section{Observations}
We obtained 1--18~GHz VLA/A-configuration data (project 19A-422; PI: Gregg Hallinan) from July 23 - October 13, 2019 for 14/26 variable radio AGN. 
For each source, the multi-band VLA observations were performed ``quasi-simultaneously" within a single scheduling block $\lesssim$ 3~hr.  This enabled a consistent and robust analysis of the radio spectral energy distribution (SED) shapes, thereby mitigating the confounding influence of intrinsic variability 
on timescales longer than a few hours.  
Further details will be presented in a forthcoming paper (Nyland et al, in prep.).  

\begin{figure}[t!]
\begin{center}
\includegraphics[clip=true, trim=0.25cm 0.72cm 0.35cm 0cm, width=0.325
\textwidth]{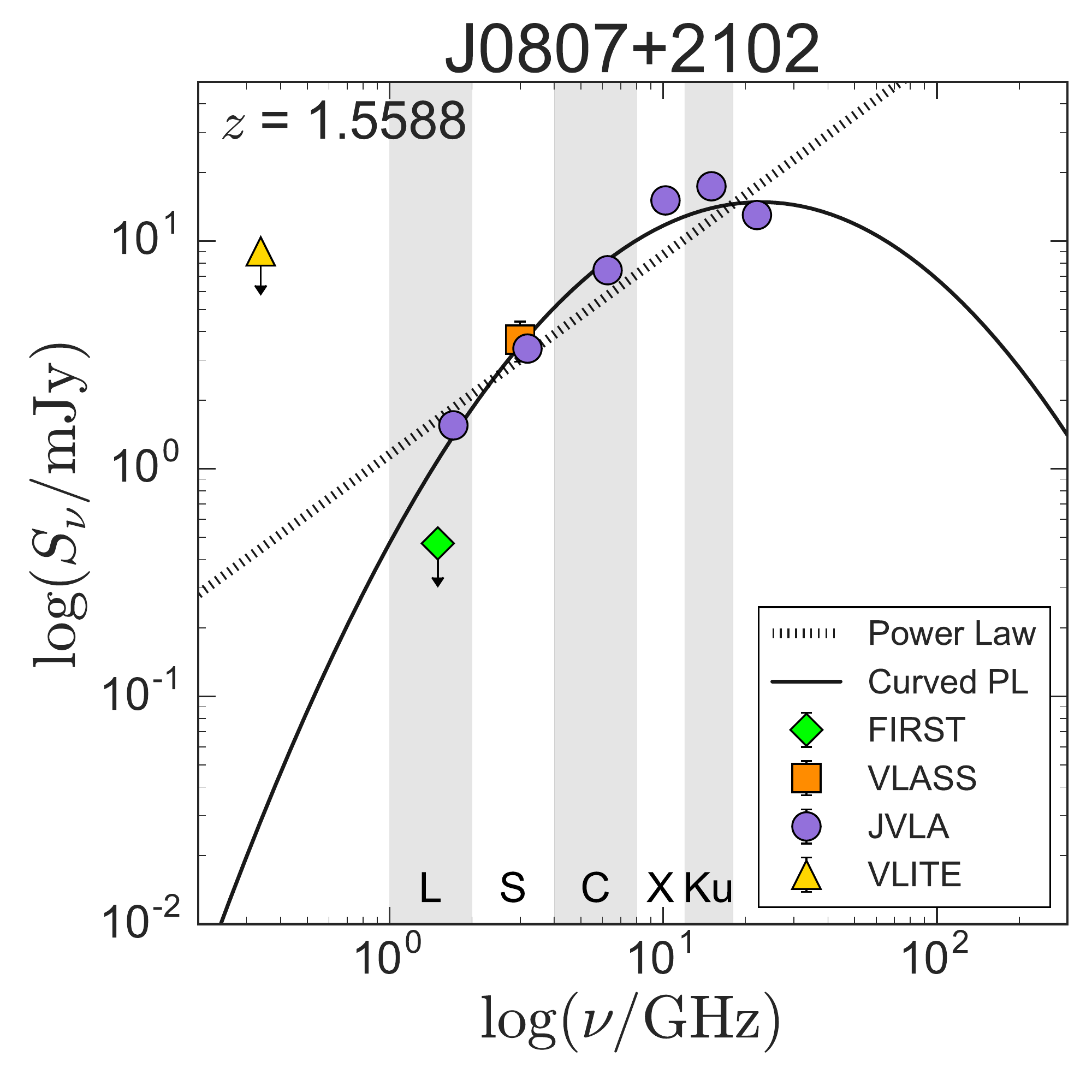} 
\includegraphics[clip=true, trim=0.25cm 0.72cm 0.35cm 0cm, width=0.325
\textwidth]{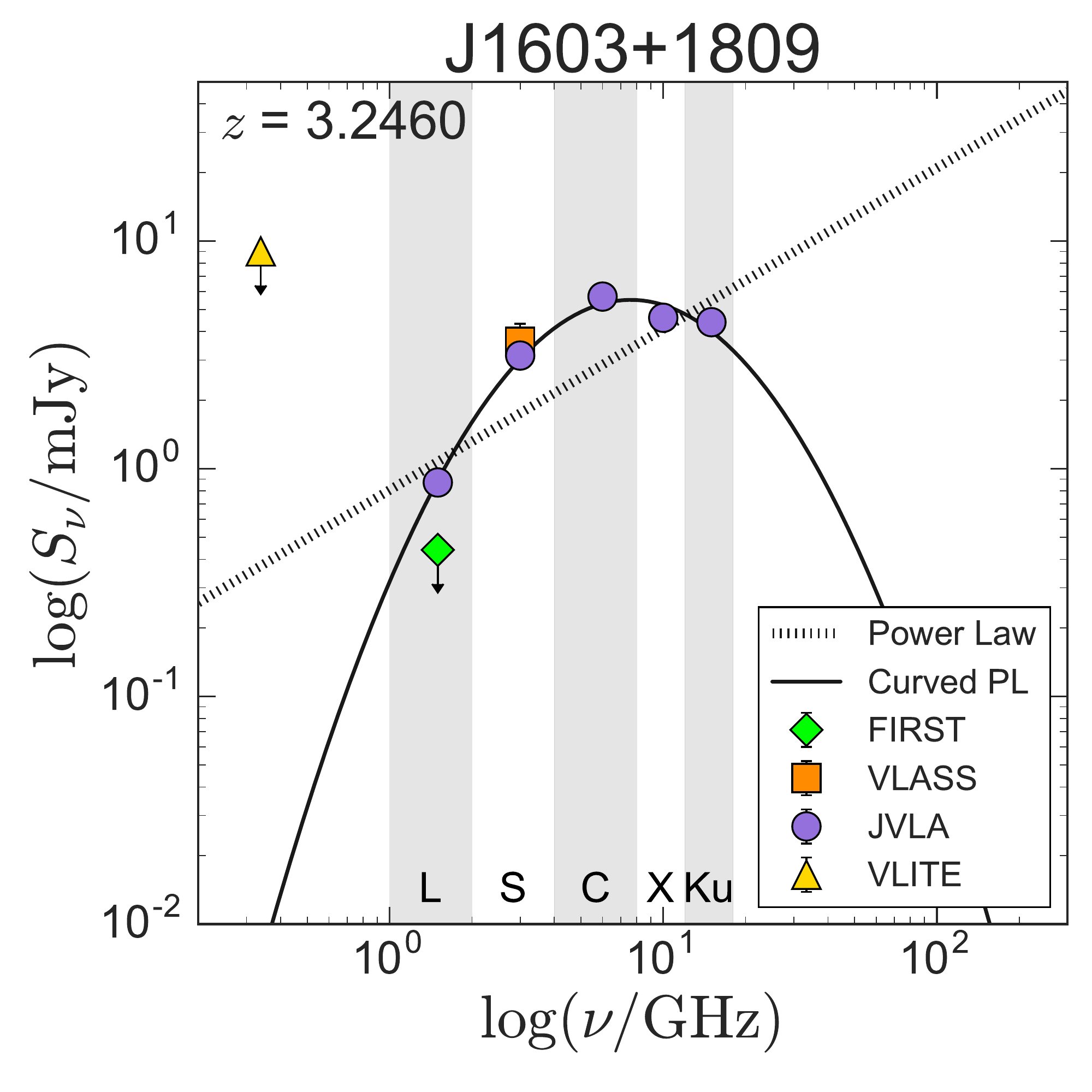} 
\includegraphics[clip=true, trim=0.25cm 0.72cm 0.35cm 0cm, width=0.325
\textwidth]{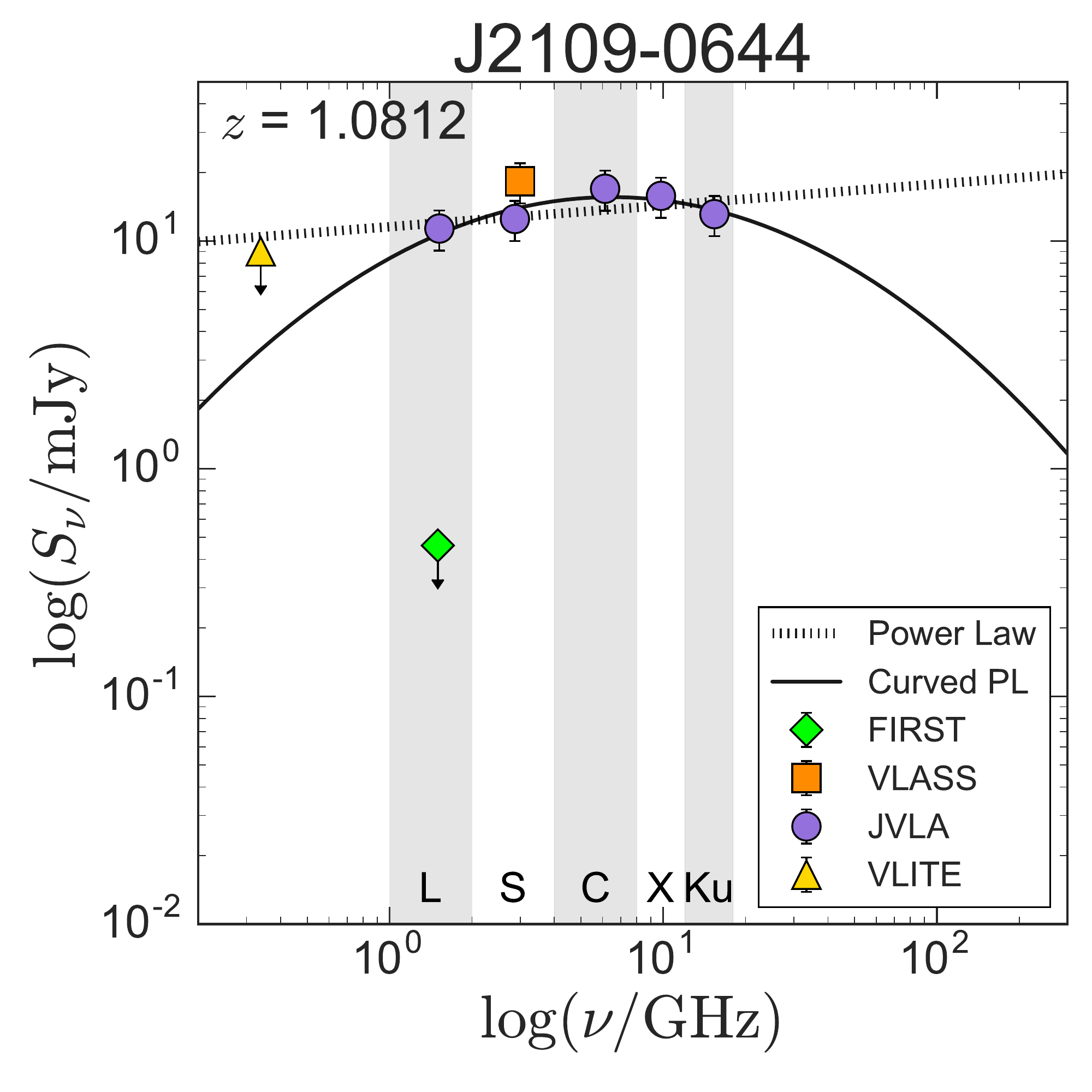} 

\caption{Example radio SEDs showing our new quasi-simultaneous VLA data (purple circles), the VLASS detections (orange squares), and the FIRST 3$\sigma$ upper limits (green diamonds).  
The yellow triangles denote 340~MHz upper limits from the commensal VLA Low-band Ionosphere and Transient Experiment (VLITE; \cite[Clarke et al. 2016]{clarke+16}; \cite[Polisensky et al. 2016]{polisensky+16}) data that were recorded during our new VLA observations.  Two model fits to our quasi-simultaneous new VLA data are shown: 1) a standard non-thermal power-model and 2) a curved power-law model.} 
\label{radio_SEDs}
\end{center}
\end{figure}

{\underline{\it Variability Constraints}}. 
Comparison between VLASS 
and our follow-up VLA $S$-band data taken a few months later reveals good agreement within the current $\sim$20\% flux uncertainties of the VLASS quicklook images (\cite[Lacy et al. 2020]{lacy+20}).  This suggests a typical variability timescale $>$ a few months. 
At $L$-band, the current peak flux values range from 1--11~mJy, which corresponds to large variability amplitudes of $\sim2-20\times$ in the 1--2 decades since FIRST.  
To put this into context, blazar variability amplitudes in the cm-wave regime are typically at the 10-20\% level (\cite[Pietka et al. 2015]{pietka+15}).

{\underline{\it Broadband Radio SEDs}}. 
In Fig.~\ref{radio_SEDs}, we show examples of the broadband radio SEDs.  We perform a least squares fit to the quasi-simultaneous data using two basic synchrotron emission models (\cite[Callingham et al. 2017]{callingham+17}, and references therein): 1) a standard non-thermal power-law model of the form $S_{\nu} = a \nu^{\alpha}$, where $a$ represents the amplitude and $\alpha$ is the spectral index,  and 2) a curved power-law model defined as $S_{\nu} = a \nu^{\alpha}e^{q(\ln \nu)^2}$, where $q$ represents the degree of spectral curvature (i.e. the breadth of the model at the peak flux value) and is defined by $\nu_{\rm peak} = e^{-\alpha/2q}$, where $\nu_{\rm peak}$ is the turnover frequency.  The presence of substantial spectral curvature typically defined as $|q| \geq 0.2$ (\cite[Duffy et al. 2012]{duffy+12}).  Our sources have significant spectral curvature ($0.18 < |q| < 1.09$) with turnover frequencies of $\sim$6.6~GHz (2.5--22.7~GHz). The quasi-simultaneous VLA data thus demonstrate that the SEDs of our sources are best modeled by curved power-law fits.  

\begin{figure}[t!]
\centering
\includegraphics[clip=true, trim=0cm 0.5cm 0cm 0cm, width=\textwidth]{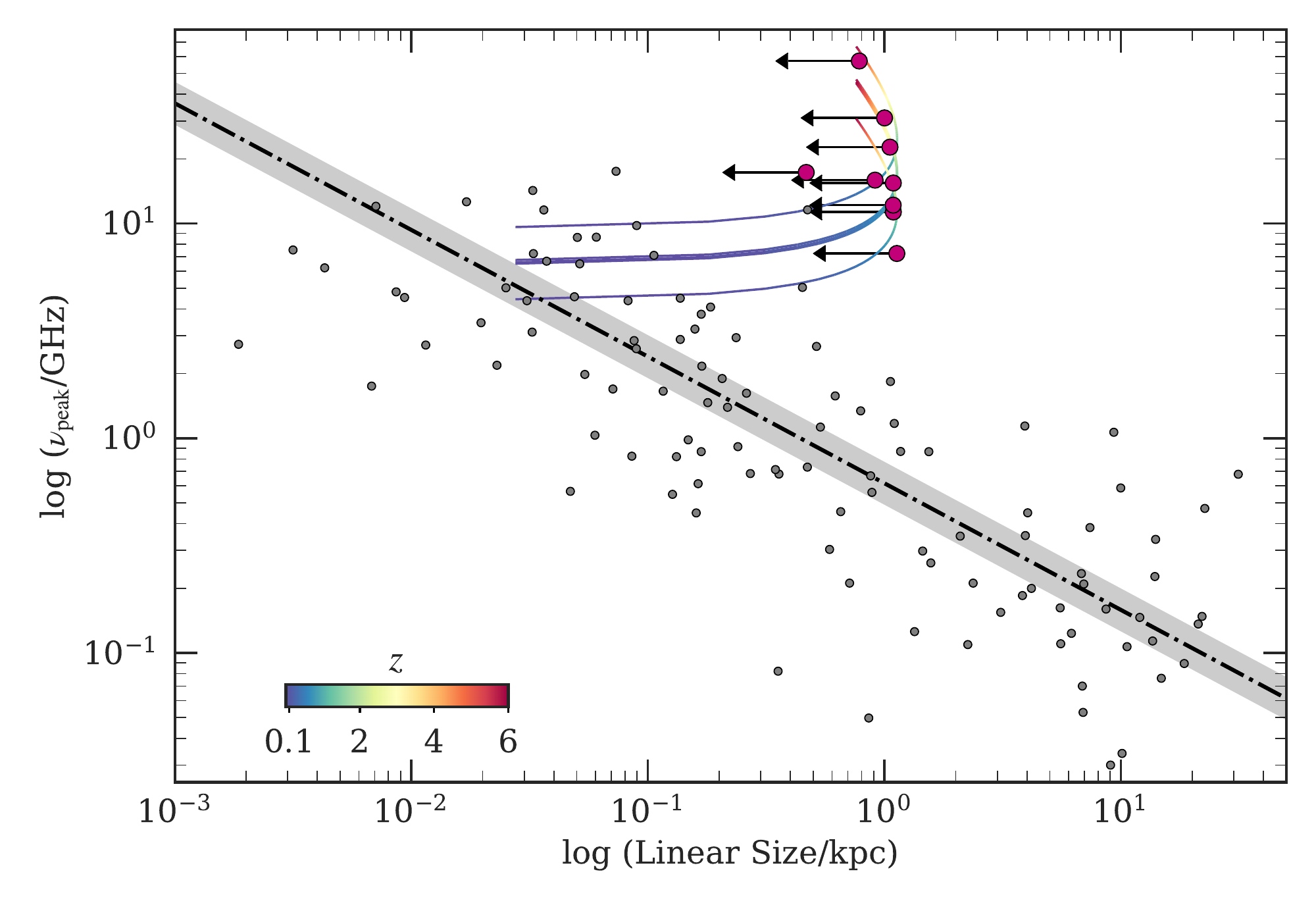}
\caption{Spectral turnover as a function of linear size. The small gray circles are literature measurements of young radio AGN (\cite[Jeyakumar 2016]{jeyakumar+16}).  The black dot-dashed line shows the empirical fit to the turnover-size relation (\cite[O'Dea 1998]{odea+98}), and the dark gray shaded region indicates the uncertainty.  Linear source size upper limits are shown by the large purple circles.  For sources lacking redshifts, the linear size upper limits are shown over a range of possible redshifts as indicated by the rainbow-colored arcs.}
\label{fig:LS_turnover}
\end{figure}

{\underline{\it Intrinsic Sizes}}. 
Our sources have compact radio emission over the full frequency range of our VLA study, placing an upper limit to their intrinsic sizes of $\theta_{\rm max} < 0.1^{\prime \prime}$ ($<1$~kpc).  
We show the turnover-size relation in Fig.~\ref{fig:LS_turnover}.  
This relationship arises from the expansion and subsequent energy loss of young jets, which causes the peak frequency (or turnover) of the relativistic electron energy distribution to shift to lower frequencies. 
Assuming our sources follow the turnover-size relation, we can obtain rough estimates of their sizes and ages.  For example, the source J0807+2102 at $z = 1.5588$ has a turnover frequency of $\sim$22.7~GHz ($\sim$58.1~GHz in the rest frame).  
We thus estimate an intrinsic jet extent of $<$1~pc and (assuming a jet speed of $0.1c$) an age of $\lesssim$30~yr for this source.    


\section{Origin of the Radio Variability?}
{\underline{\it Extrinsic vs. Intrinsic Variability}}. 
The radio SED shapes, variability timescale and amplitude constraints, high radio luminosities, 
and high SMBH masses rule-out variability driven by extrinsic propagation effects or transient phenomena (including on-axis GRBs, radio supernovae afterglows, and tidal disruption events).  
We thus conclude that the increase in radio flux is caused by \emph{intrinsic} AGN variability. 
Intrinsic radio AGN variability may arise from blazar-like behavior (e.g.\ shocks propagating along the jet due to magnetic field turbulence; \cite[Marscher \& Gear 1985]{marscher+85}), the rapid re-orientation of a compact jet towards our line of sight (e.g.\ \cite[Bodo et al.\ 2013]{bodo+13}), and newborn radio jets that have been recently launched (possibly following a state transition of the accreting SMBH; \cite[Wojtowicz et al. 2020]{wojtowicz+20}).  

{\underline{\it Young Radio Jets}}. 
Given the 
radio properties of our sources (in particular variability amplitude and timescale constraints), we favor the jet youth scenario in which the radio variability is the result of parsec-scale jets that were launched in the last 1--2 decades.   
Confirmation of the jet youth scenario will require continued monitoring of the evolution of radio fluxes and SED shapes, as well as tighter size constraints from higher-resolution radio observations. 

Young radio AGN, such as Gigahertz peaked spectrum (GPS) sources, are characterized by compact morphologies and inverted radio SEDs below their turnover frequencies, which are typically in the GHz regime (\cite[O'Dea 1998]{odea+98}), consistent with the morphologies and radio SEDs of our sources. After a jet is launched, models predict a rapid increase in luminosity ($P_{\rm radio} \sim t^{2/5}$) as the dominant energy loss mechanism transitions from adiabatic to synchrotron losses (\cite[An 2012]{an+12}), making the identification of young radio AGN in VLASS that have emerged in the time since FIRST ($\sim20$~yr) plausible.  The identification of such young radio AGN is not unprecedented; the youngest known sources have kinematic ages as low as 20~yr (\cite[Gugliucci et al. 2005]{gugliucci+05}).  For a nascent radio jet that has been triggered within the last 20~yr, the model of (\cite[An 2012]{an+12}) suggests a $>3\times$ increase in radio luminosity, which is consistent with the observed radio brightening at 1.4~GHz between FIRST and our 2019 $L$-band observations.  

\section{Discussion}
{\underline{\it Radio-changing-state Quasars?}} 
While our sources would have previously been classified as radio-quiet based on their upper limits in FIRST, VLASS has revealed that they are now consistent with radio-loud quasars (e.g. \cite[Kellerman et al. 2016]{kellerman+16}).  
Other multi-epoch radio surveys have recently identified AGN with similar radio variability amplitudes and timescales and concluded that jets with lifetimes of $10^{4-5}$~yr may be common (\cite[Jarvis et al. 2019]{jarvis+19}, \cite[Wolowska et al. 2017]{wolowska+17}). 
If this is indeed the case, jet-ISM feedback may play a more important role in the regulation of SMBH growth and star formation than 
typically assumed.  Large statistical studies of compact radio AGN will ultimately enable key improvements to prescriptions for sub-grid models of AGN feedback in cosmological simulations.  

{\underline{\it Future Prospects.}} 
As part of our continued effort to study AGN that have recently transitioned from radio-quiet to radio-loud, new multi-band radio observations with the Very Long Baseline Array, which will provide milliarcsecond-scale resolution, are currently in progress.  Over the next several years, the completion of the remaining epochs of VLASS, as well as future multi-epoch, high-resolution studies with prospective instruments such as the next-generation Very Large Array (\cite[Nyland et al. 2018]{nyland+18}), will enable major advancements in our understanding of the triggering of radio jets and their connection to galaxy evolution.  

\end{document}